\documentstyle[epsfig,twocolumn]{mn}

\newcommand{\delphi}{{\varpi}}
\newcommand{\rg}{{r_{_{\rm{S}}}}}
\newcommand{\dg}{{^{\rm{o}}}}
\def\lb#1{{\protect\linebreak[#1]}}

\title[A cloud model of active galactic nuclei]{A cloud model of active
 galactic nuclei: the iron K${\protect\bmath{\alpha}}$ line diagnostics}
\author[V. Karas et al.]
 {V.~Karas,$\!^1$\thanks{E-mail: 
   vladimir.karas@mff.cuni.cz (VK);
   bcz@camk.edu.pl (BC); 
   arnaud.abrassart@obspm.fr (AA); 
   marek@fy.chalmers.se (MAA)}
  B.~Czerny,$\!^2$
  A.~Abrassart$^3$ and
  M.\,A.~Abramowicz$^4$\\ 
 $^1$Astronomical Institute, Charles University Prague, 
  V~Hole\v{s}ovi\v{c}k\'ach~2, CZ--180\,00~Praha, Czech Republic\\
 $^2$Nicolaus Copernicus Astronomical Center, Bartycka~18,
  PL--00\,716~Warsaw, Poland\\
 $^3$DAEC, Observatoire de Paris, Section de Meudon, F-92\,195~Meudon,
 France\\
 $^4$Institute for Theoretical Physics, G\"oteborg University and
 Chalmers University of Technology, S--412\,96~G\"oteborg, Sweden}

\date{Accepted ................. 2000; Received ................. 1999}
\pagerange{\pageref{firstpage}--\pageref{lastpage}}

\begin{document}
\label{firstpage}
\maketitle

\begin{abstract}
The origin and the profile of the iron K$\alpha$ line from active
galactic nuclei is being studied. The model with a quasi-spherical
distribution of clouds around a compact galactic nucleus is explored in
which the intrinsic spectral feature around 6--7\,keV is substantially
affected by Comptonization of primary X-rays. The predicted spectral
profiles are influenced mainly by motion of the clouds in strong
gravitational field of the centre, and by orientation of the source with
respect to the observer. The case of the Seyfert galaxy MCG--6-30-15 is
examined in more detail and free parameters of the model are determined.
A robust result is obtained: acceptable fits to the observed profile
correspond to the spherical distribution of the clouds with strong
self-obscuration effects and with rather high value of ionization
parameter. This is a different situation from that considered in
the model of a cold illuminated disc. Nevertheless, geometrically flat
(disc-type) configurations are not rejected.

\end{abstract}


\section{Introduction}
An ever-increasing accuracy of the determination of the observed shape
of the iron K${\alpha}$ line in the Seyfert~1 galaxy MCG--6-30-15, and
the evidence of the broad and skewed profile (Tanaka et al.\ 1995;
Iwasawa et al.\ 1996) have led to wide acceptance of the model with an
accreting black hole in the nucleus, and it offered an unprecedented
opportunity to explore directly the pattern of the accretion flow onto
the central hole in active galactic nuclei (AGN; for recent and detailed
expositions of the subject, see Peterson 1997; Krolik 1999). The iron
reflection features and remarkable variability patterns have been
reported in many other AGN, galactic black-hole candidates, and similar
objects. 

It is the main aim of the present paper to examine relationship
between the form of motion of the gaseous material in a galactic nucleus
and the resulting profile of the spectral features. We concentrate
ourselves on the question whether current observational evidence can be
explained in terms of individual clouds with spherical or almost
spherical (rather than disc-type) orbital motion around the centre, or
if strongly flattened distribution of the clouds is preferred resembling
a disc or a ring (different models of the line formation in AGN were
reviewed by Netzer 1990). This topic has far-reaching consequences for
the unification scheme of AGN (Antonucci 1993). Although the assumptions
of the disc geometry and of strictly planar bulk motion of the gaseous
material are relaxed in the present paper, the presence of a compact
supermassive accreting nucleus remains crucial for explaining the
observed spectral features.

Broad iron features were expected in X-ray spectra on the basis of the
model in which the iron line is formed on the surface of a geometrically
thin, optically thick and relatively cold medium after irradiation by a
primary source (Fabian et al.\ 1989). Subsequent detailed fits to
observational data confirmed that the line profile is in agreement with
the predictions of the accretion disc model in which the intrinsically
narrow line, emitted at energy 6.4\,keV, is gravitationally shifted and
Doppler broadened/boosted due to the disc orbital motion. Comparisons of
the model with the data allow one to constrain the disc inclination angle,
the range of disc radii contributing to the emission, and the radial
dependence of the incident X-ray flux, although
there are various uncertainties if astrophysically more realistic models
of accretion flows are introduced, and if the lack of resolution and the
noise in available data are taken into account; cf.\ Fabian et al.\
(1989); Tanaka et al.\ (1995); Weaver \& Reynolds (1998) for the
case of iron line diagnostics, and Rokaki \& Boisson (1999)
for application to UV continuum and H$\beta$ emission line.

The expected amplitude of the line profiles and the characteristic form
of continuum from X-ray illuminated accretion flows were examined,
taking into account combination of effects due to high orbital
velocities and strong gravity near the nucleus (George \& Fabian 1991;
Matt, Perola \& Stella 1993). It has been argued with various levels of
refinement that parameters of the central black hole (especially its
angular momentum) can be inferred from disc-line spectra, assuming that
they are sensitive to the radius of the innermost stable orbit whose
imprint is visible in radiation of the accretion flow (Laor 1991;
Iwasawa et al.\ 1996; Dabrowski et al.\ 1997; Pariev \& Bromley 1998). 

Reynolds \& Begelman (1997) pointed out, however, that the model
is not that sensitive to the value of the spin of the hole if the
contribution to the line is allowed also from matter inspiralling below
the marginally stable orbit, $r_{\rm{ms}}$. Such a possibility appears
as perfectly consistent with high efficiency of X-ray radiation if the
adopted geometrical depth of the stream is small, in agreement with
actual computations of the flow properties below $r_{\rm{ms}}$
(Muchotrzeb \& Paczy\'{n}ski 1982). 

In the case of accretion-disc
geometry, and with different assumptions about the X-ray illumination
and reprocessing, the predicted spectra (line plus continuum) have been
calculated by several authors (recently Young, Ross \& Fabian 1998;
Martocchia, Karas \& Matt 2000; see further references therein).
The same scheme, coupling the shape and the
amplitude of the line with the shape and the amplitude of the Compton
reflection component, was successfully applied to galactic black-hole
candidates (\.{Z}ycki, Done \& Smith 1998; Done \& \.{Z}ycki 1999).

The line fits are consistent with the line emitted at intrinsic energy
6.4\,keV (so it comes from weakly ionized iron) while the broad shape
of the spectral feature indicates that it is formed near the innermost
part of the disc, following intense irradiation. This might be related
to the specific slope of the hard X-ray emission (R\'{o}\.{z}a\'{n}ska
et al.\ 2000; Nayakshin, Kazanas \& Kallman 1999). The blue wing of the
line is in all sources linked with the broad feature around 6.4\,keV,
requiring somewhat special interplay between model parameters. Namely,
strong constraints are imposed on the inclination angle of the disc [see
Guainazzi et al.\ (1999) for the case of MCG--6-30-15; Wang, Zhou \&
Wang (1999) for NGC\,4151; Nandra et al.\ (1999) for NGC\,3516] although
contribution to the total light from various components helps to ease
this constraint to some extent.

Sulentic et al.\ (1998b) raised the problem of disagreement between the
inclination angle derived by two independent approaches: from the
iron-line model, and from H${\beta}$ and H${\alpha}$ measurements. In a
large sample of objects, the position of the line centroid at 6.4\,keV
is not consistent with the random orientation of the disc with respect to
the observer (Sulentic, Marziani \& Calvani 1998a) although this problem
is weakened by the fact that the Balmer lines do not necessarily have to
come from the outer parts of the disc, as assumed in the paper. In
particular, the mean inclination derived for a sample of Seyfert~2
galaxies does not differ from the mean inclination angle of Seyfert~1
galaxies (Turner et al.\ 1998), which disagrees with the widely
preferred unification scheme (however, see Weaver \& Reynolds 1998).
Also, the outer disc radius comes out lowish, of the order of ten
Schwarzschild radii ($\rg\dot{=}2.95\times10^5\,M/M_{\odot}$\,cm in
terms of the central black-hole mass $M$).
The problem of this apparent disagreement in inclination angles
can be weakened by the fact that the Balmer lines do not necessarily have 
to come from the outer parts of the disc, as assumed in the paper,
and there is also a contribution from the
narrow unresolved component (expected to arise in the dusty/molecular torus;
Krolik, Madau \& \.Zycki 1994) which has not been properly acounted.

Certain doubts concerning the disc model for the iron-line production
revived the interest in the alternative explanation of the line profile
(Czerny, Zbyszewska \& Raine 1991). Misra \& Sutaria (1999) assume that
the line is produced by Compton scattering of the line photons in warm,
Thomson thick material. They show that such a model fits the data
equally well as the disc model, but the approach may look rather ad hoc
because the origin and location of the Comptonizing medium have not been
addressed in their paper, neither the source of the X/UV incident
continuum that is a necessary ingredient of this model. A better
motivated approach was adopted by Abrassart (2000a; see further
references cited therein) who explored spectral properties within the
frame of the clouds model of accretion onto a black hole where both the
line formation and Comptonization occur in the same medium. However,
his results did not incorporate the global effects (gravitational
redshift, clouds motion, etc). Hereafter we will argue that these
effects cannot be ignored.

In the present paper we discuss the possibility of explaining the Fe
K${\alpha}$ line within the frame of the model in which the innermost
part of the disc is disrupted due to disc instabilities. Part of the
disc material forms optically thick cold clouds, while another fraction
heats up to high temperatures acting as a source of X-rays. The clouds
are not confined to the disc equatorial plane, and they form a layer
covering a significant portion of the sky from the point of view of the
central X-ray source (Collin-Souffrin et al.\ 1996). In this model the
line profiles are determined by two concurrent effects: they arise
partially from the Comptonization within the material significantly
ionized at the surface of the clouds (the intrinsic line profile), and
partially from kinematics of the clouds distribution and from strength
of the gravitational field where the clouds persist (smearing of the
profile). In the next section we formulate a simplified model which
captures the essence of the clouds scenario and can be further developed
to a more realistic form. We show the predicted line profiles as a
function of model parameters (specified by the clouds distribution) and
the observer view angle. The intrinsic shape of the spectral feature has
been taken either as a narrow delta-type line, or a numerically computed
broad feature (corresponding to high ionization parameter $\xi$). Then
we briefly discuss how various complications (obscuration of the clouds,
non-Keplerian orbital motion) are reflected in resulting profiles
(Sec.~\ref{results}). We conclude the paper by comparison with the
{\it{ASCA}\/} data for MCG--6-30-15, and we discuss overall advantages
and the problems of this picture.

\begin{figure}
\epsfxsize=\hsize
\epsffile{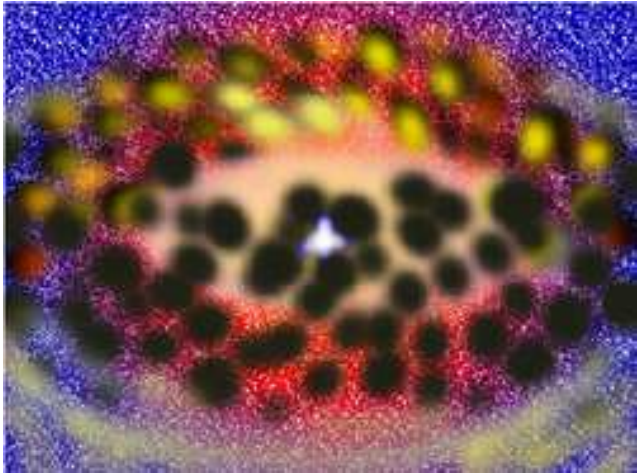}
\caption{A schematic view showing the main ingredients of the model.
Most of the clouds are distributed at some characteristic radius, their
origin being presumably in the accretion disc disrupted due to
instabilities. The clouds survive as individual entities orbiting around
the massive and compact core. They are irradiated from the centre, and
the primary X-rays are reprocessed on the clouds surfaces. From the
observer's view, dark parts of the clouds are in front of the nucleus
while radiating surfaces correspond to the clouds in the back side. The
intrinsic spectrum is determined mainly by the Comptonization, depending
sensitively on ionization parameter. Rather clean line of sight to the 
central source emerges among the clouds. The observed spectral features are
then influenced by the clouds distribution, their velocities, and
self-obscuration effects. Also the aspect angle of the observer must be
considered if the distribution is flattened rather than spherically
symmetric.
\label{fig0}}
\end{figure}

\begin{figure}
\epsfxsize=\hsize
\epsffile{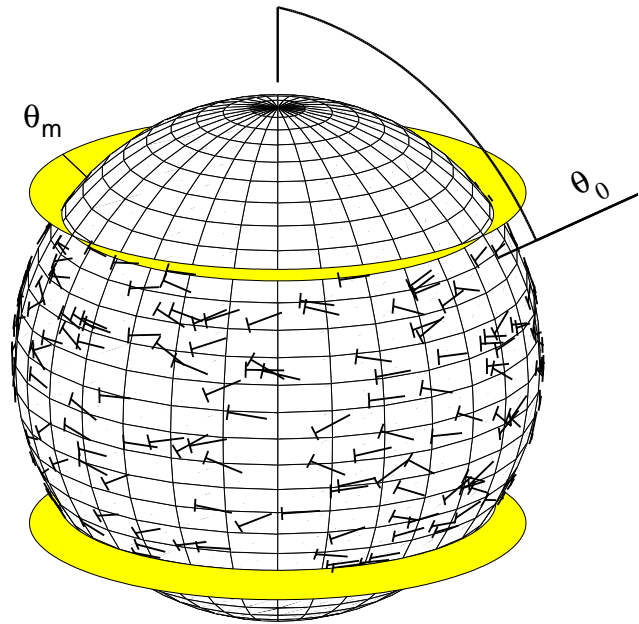}
\caption{A representation of the clouds distribution (positions and
velocities of the clouds). In this sketch we consider all the clouds
being located on the surface of a sphere. Their trajectories are
characterized by constant value of the radius $r$, and by maximum
excursion $\theta_{\rm{m}}$ from the equatorial plane. The clouds are
distributed randomly within the allowed range of latitude which also
restricts directions of the clouds velocities (indicated by short
dashes). Observer's inclination angle $\theta_0$ is also shown.
\label{fig1}}
\end{figure}

\section{The model}
We assume that the iron K${\alpha}$ line is produced by irradiated
surfaces of the clouds. The clouds could be formed from the inner
accretion disc which is eroded at a distance of a few or a few tens of
Schwarzschild radii. The basic principles of the model were outlined by
Collin-Souffrin et al.\ (1996), and the broad-band spectra following
from this model were studied by Czerny \& Dumont (1998). The expected
shape of K${\alpha}$ line was calculated with careful consideration of
X-ray reprocessing (Abrassart 2000b) but without taking into account any
kinematical and gravitational effects which must influence the observed
spectral features substantially if the clouds are to be located so close
to the central black hole. Resulting width and centroid energy of
the line were thus entirely due to the Comptonization. In those papers,
the clouds distribution was assumed to be spherical and positions of all
the clouds were fixed at a given distance from the black hole.

Here, we relax the assumption of stationary clouds forming a spherical
layer with strictly uniform distribution. Instead, the clouds follow
planar orbits departing from the equatorial plane of the disc
(Figure~\ref{fig0}). One of the appealing advantages of this scheme is
the naturally modeled variability of X-ray sources due to obscuration
events (Abrassart \& Czerny 2000). In the present paper it is assumed
that excursions of the clouds from equatorial plane are limited by the
maximum angle of deflection, say $\theta_{\rm{m}}$. This angle, and
radius of the clouds belt, $r$, stand as two free parameters in the
simplest version of the model. The assumed distribution accommodates
both the case developed by Collin-Souffrin et al.\ (1996), and the case
of a narrow ring (Gerbal \& Pelat 1981) in the limiting values of
$\theta_{\rm{m}}$. In principle, our model can accomodate a system of clouds
on both prograde and retrograde orbits with respect to the disc, but 
an intermediate value of the clouds departure from
the equatorial plane appears as
most favourable, corresponding to quasi-spherical or somewhat 
flattened system of the clouds
following more or less corotating trajectories. This case
allows for quite high covering factor (necessary in the model which
relies on multiple scattering of radiation among the clouds) 
while reducing the rate of collisions
(which otherwise tend to destroy the the clouds).

The present model
does not require all the individual clouds to preserve their identity
for many orbital periods or even indefinitely.
Although it may appear easier in this scheme to think of the clouds as
separate entities, the predicted spectrum will not
change if the clouds have a finite and very limited lifetime,
assuming they are continously replaced by newly created ones.
Each of the clouds can be removed from the system after, say,
one revolution, either due to collisional destruction with other clouds
or by heating up to the state when it does not contribute to the line
spectrum and possibly evaporates. In the other words, the spectral
features would be expected more pronounced during the initial phase of
the cloud's history before reaching the equilibrium.

Radiation of the clouds surface is characterized by intrinsic emissivity
which is assumed constant and isotropical in the local comoving (rest)
frame of each cloud. Since the distribution is not spherically symmetric
in general, the observed spectrum depends on inclination angle of the
observer. We neglect the contribution to the line coming possibly from
the outer, fairly smooth part of the disc. The intrinsic line profiles
of the clouds come out somewhat different from those which are usually
adopted in the disc-line model (cf.\ Sec.~\ref{results}), and the random
motion of the clouds must be also taken into account in calculations of
predicted spectra.

\subsection{Kinematical and gravitational effects}
The basic observational properties relevant for this paper, i.e.\ the
observed profiles of spectral features (their centroid energies and
widths in particular), can be roughly predicted by a simplified model in
which the clouds distribution is specified by a small number of
phenomenological parameters. Notice, however, that one still needs to
solve for the radiation reprocessing in order to obtain the intrinsic
line profiles.

In the simplest version of the model, the observer is located at
inclination $\theta_0$ far from the source ($0\leq\theta_0\leq\pi/2$),
while all the clouds are distributed at the same radial distance and
they move with the corresponding Keplerian orbital velocity,
$v_{\rm{k}}(r)$. Gravitational field of the central body is spherically
symmetric and described by the Newton law. We introduce Schwarzschild
radius ${\rg}=2GM/c^2$ as a convenient length-scale and we use
geometrized units with $G=c=1$ hereafter.
 
The clouds are distributed randomly within a band of spherical
latitudes: $\pi/2-\theta_{\rm{m}}\leq\theta\leq\pi/2+\theta_{\rm{m}}$
(Figure~\ref{fig1}). Spherical coordinates $r$, $\theta$, $\phi$ are
employed with the disc plane at $\theta=\pi/2$. Parameter
$\theta_{\rm{m}}$ reflects the dynamics of the process of clouds
formation and of their interaction with surrounding environment. If the
disc disruption proceeds rapidly, and if the newly formed clouds are
efficiently captured by some outflowing plasma and/or accelerated by the
radiation pressure, then the clouds distribution can be approximately
characterized by $r={\rm{const}}$ with relatively large
$\theta_{\rm{m}}$. On the other hand, if the clouds acceleration is only
moderate in the direction perpendicular to the disc plane, and if their
subsequent evaporation is fast in comparison with the timescale of
latitudinal motion, then the distribution is again well characterized by
constant radius but rather small $\theta_{\rm{m}}$.

\begin{figure}
\epsfxsize=\hsize
\epsffile{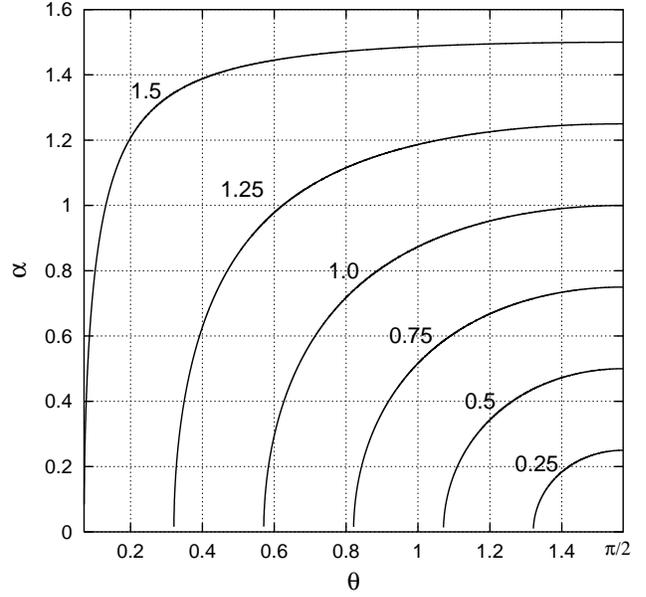}
\caption{The range of allowed values of angle $\alpha$ constraining the
velocity direction of the clouds as a function of $\theta$ [cf.\
eq.~(\protect\ref{alpha})]. The curves are labeled with the
corresponding latitudinal boundary $\theta_{\rm{m}}$ of the clouds belt,
which adopts values from $\theta_{\rm{m}}=\pi/2$ (strictly spherical
distribution) down to $\theta_{\rm{m}}=0$ (the equatorial ring).
\label{fig2}}
\end{figure}

Velocity of the clouds points along $r={\rm{const}}$ surface and
its direction defines the angle $\alpha$:
\begin{equation}
\bmath{v}=v_{\rm{k}}(\cos\alpha\,\bmath{e_\phi}-\sin\alpha\,\bmath{e_\theta}).
\label{alpha}
\end{equation}
The range of possible values of $\alpha$ is determined by the angular
width $\theta_{\rm{m}}$ of the clouds distribution. Straightforward
trigonometry gives relation for the maximum $\alpha$ of the clouds
velocities,
\begin{eqnarray}
 \sin\alpha&=&\pm
 \left(\cos\theta\cot\theta\cot\theta_{\rm{m}}\cos\theta_{\rm{m}}-
 \sin\theta\sin\theta_{\rm{m}}\right)
 \nonumber \\
 &&\times\left(1-\cos^2\theta\sin^{-2}\theta_{\rm{m}}\right)^{1/2}
 \nonumber \\
 &&\mp\cos^2\theta\sin^{-1}\theta_{\rm{m}}
 \left(1-\cot^2\theta\cot^2\theta_{\rm{m}}\right)^{1/2}.
 \label{alpha2}
\end{eqnarray}
This relation is shown in Figure~\ref{fig2} where $\theta_{\rm{m}}$
stands as parameter of the curves $\alpha(\theta;\theta_{\rm{m}})$. For
example: one can read in the graph that latitudinal motion of the clouds
has turning points at $\alpha(\pi/2-\theta_{\rm{m}};\theta_{\rm{m}})=0$
where trajectories are parallel to the disc plane (i.e.\
$\bmath{v}\parallel\bmath{e_\phi}$). The upper/lower signs in
eq.~(\ref{alpha2}) correspond to the descending/ascending parts of the
trajectory, respectively. We recall that for
$\theta_{\rm{m}}\rightarrow0$ the clouds remain in the equatorial plane
(ring-like distribution), while for $\theta_{\rm{m}}=\pi/2$ the clouds
are spread uniformly over the whole sphere (with random distribution).

The observed radiation of a cloud is determined by special-relativity
effects which influence photon energy $E$ and intensity $I(E)$ in usual
manner (Cunningham \& Bardeen 1973): $I(E)=I_{\rm{e}}(E/g)\,g^3$ where
$I_{\rm{e}}$ is the emitted intensity, and $g$ is the redshift factor
defining the change of energy. We adopt standard notation, similar to
Ohna et al.\ (1995) who studied a related problem with application to
fast winds. In our case, however, it is the inner part of the clouds
surfaces which is hot enough to emit the X-rays. The remote part of the
clouds (with respect to the central source) is not subject to primary
irradiation, and it emits almost no X-rays at all. As a consequence of
self-obscuration, X-rays are visible only from the clouds with
appropriate orientation, and their radiating area is further reduced on
the rim of the image by projection effect. In addition, mutual eclipses
of the clouds must be considered. The resulting obscuration is enhanced
in the case of clouds which substantially overlap each other; we will
characterize this effect by another free parameter [$\omega$; cf.\
eq.~(\ref{obscuration}) below].

\subsubsection{The case of negligible obscuration}
First we ignore entirely the obscuration of the clouds by other clouds.
We assume that the clouds radiate isotropically in their local comoving
frame, and they are of spherical shape with an infinitesimally small
radius (i.e.\ much less than ${\rg}$). In this case all the clouds from
the whole hemisphere contribute to observed X-rays.

The redshift factor $g$ depends on the angle between velocity of the
cloud and the unit vector $\bmath{s}$ towards the observer,
\begin{equation}
 g=L^{-1}\gamma\,(1-\bmath{v.s}).
 \label{g}
\end{equation}
Here, $\gamma=1/\sqrt{1-v^2}$ is the Lorentz factor, and
$\bmath{v.s} = v_{\rm{k}}\cos\theta_0\sin\alpha\sin\theta
-v_{\rm{k}}\sin\theta_0(\cos\alpha\sin\phi
+\sin\alpha\cos\theta\cos\phi)$.
Integration of light over the clouds distribution yields the total
observed radiation flux. Effects of general relativity are accounted by
the gravitational redshift term $L=\sqrt{1-{\rg}/r}$, but gravitational
lensing has been ignored here for simplification (this might have some
effect on the light from clouds at the upper conjunction).

Since we deal with the clouds that are irradiated by the central source,
one expects the local emissivity to be depending on the angle between
the direction towards center and the normal to the cloud's surface. In
such a case, by a simple geometrical argument, the observed line profile
of the whole axisymmetric collection of the clouds is the same as for
the case of clouds radiating truly isotropically. This conclusion does
not hold, however, if partial and non-axisymmetric self-obscuration of
the clouds is involved.

\subsubsection{Partial obscuration of the clouds}

\begin{figure}
\epsfxsize=\hsize
\epsffile{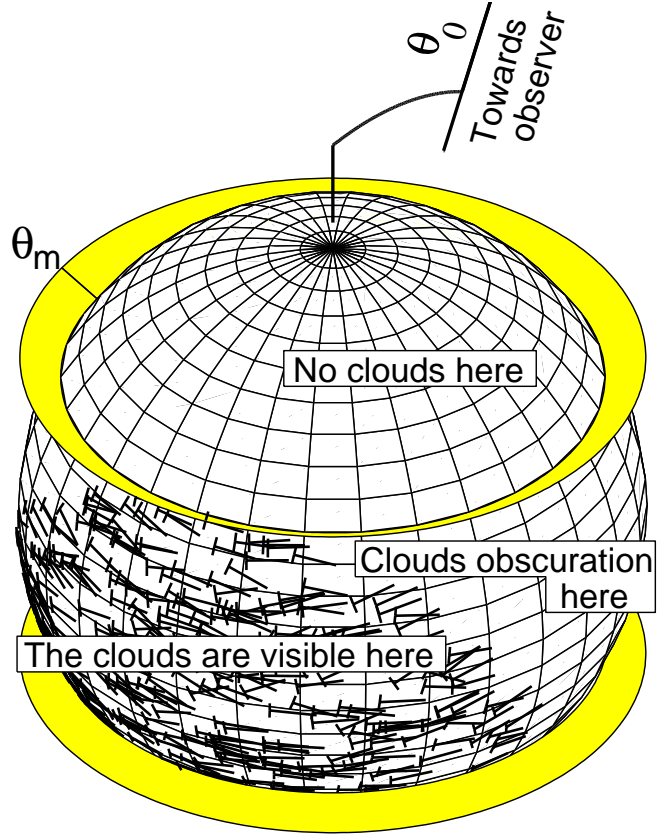}
\caption{A view of the clouds distribution in the case of non-negligible
obscuration. Positions and velocities of those clouds are shown which
fulfil conditions of visibility (\protect\ref{obscuration}). In
comparison with Fig.~\protect\ref{fig1}, only a fraction of the clouds
contribute to the observed spectrum: those moving more or less
perpendicularly to the observer's line of sight. Direction towards the
observer is shown with inclination angle $\theta_0$. Different regions
are indicated on the sphere according to visibility of the clouds. The
clouds do not reach high latitudes, and they enter in dead angle near the
rim of the projected image. This figure corresponds to parameters
$\theta_{\rm{m}}=30\dg$, $\theta_0=60\dg$, $\omega=60\dg$ in our
notation.
\label{fig3}}
\end{figure}

The presence of a large number of clouds (typically $N\approx10^3$)
enhances the chance of mutual obscuration among them. This effect cannot
be ignored in this model because the observer receives X-rays only from
those parts of the clouds which are on the inner side of the
$r={\rm{const}}$ surface, and not too close to the edge of the image (as
projected onto the observer image plane). It can be included in the
model by introducing another parameter, say $\omega$, characterizing the
maximum angle between the line of sight and the edge of the clouds
distribution: $\bmath{s.e_{r}}=\cos(\pi-\omega)$. In other words,
outgoing rays must pass from interior hot surfaces of the clouds through
empty holes in between them before they can finally escape towards the
observer. This is impossible near the projected rim of the sphere.
Therefore in calculations of the observed radiation, only those clouds
are considered which, in addition to the conditions described in the
previous paragraph, satisfy also relation (Figure~\ref{fig3})
\begin{equation}
\sin\theta_0\sin\theta\cos\phi+\cos\theta_0\cos\theta>-\cos\omega.
\label{obscuration}
\end{equation}

Let us remark that the angle $\omega$ can be related to the covering
factor $f_{\rm{c}}$ (defined as the fractional area of the sky subtended
by the clouds as viewed from the centre of the source). The covering
factor $f_{\rm{c}}$ can be expressed in terms of the clouds typical
diameter $d$, typical distance $l$ between the clouds, and their number
$N$: $f_{\rm{c}}=N(d/r)^2$, with the obscuration constraint
$d{\approx}l\cos\omega$ for spherical clouds filling the whole
$r={\rm{const}}$ surface ($f_{\rm{c}}$ obviously decreases with
$\theta_{\rm{m}}$ decreasing). The exact description of obscuration
would require a specification of the clouds shapes, and it can be carried
out only numerically by generating random clouds distributions with the
constraints imposed by the visibility conditions, checking the line of
sight in each case. Furthermore, $d$ and $l$ are linked to each other
via $N$, but parameter $\omega$ has direct geometrical meaning in our
model, and it appears thus convenient for the present discussion. In
principle, the same value of the phenomenological parameter $\omega$ can
correspond to different physical models of the clouds origin,
their geometrical configuration and other details.

\subsection{The intrinsic line profile}
\label{secemis}
                          
\begin{figure}
\epsfxsize=\hsize
\epsffile{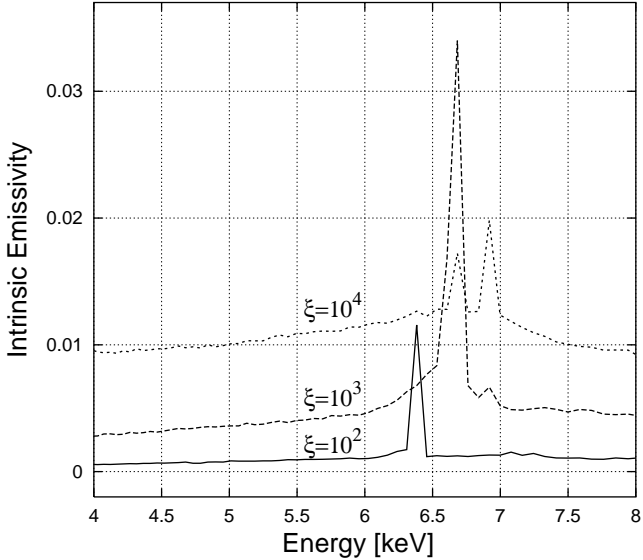}
\caption{The intrinsic emissivity profile (arbitrary units) around the
6--7\,keV spectral feature, as predicted by the Comptonization model for
hot parts of the clouds surfaces. The three curves correspond to
different values of the ionization parameter $\xi$ (Abrassart 2000b).
\label{figline}}
\end{figure}

\begin{figure}
\epsfxsize=\hsize
\epsffile{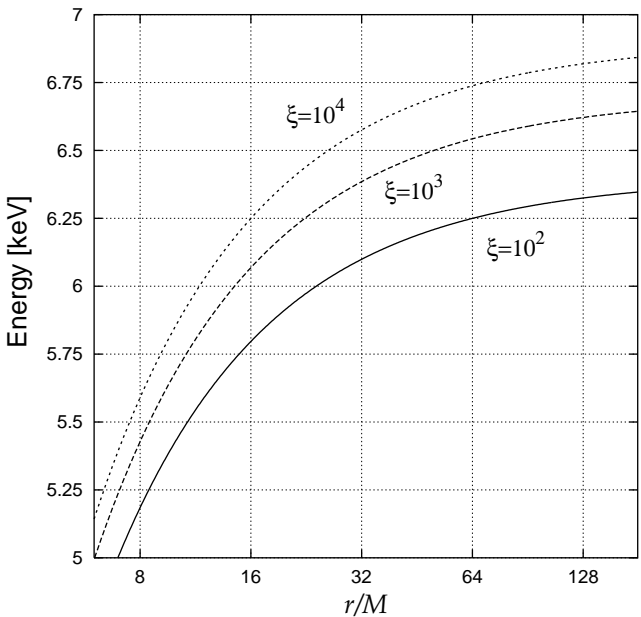}
\caption{Relation between radius (logarithmic scale) and centroid energy
of the line features dominating the intrinsic spectrum, after
multiplication by that part of the redshift factor from
eq.~(\protect\ref{g}) which depends on $r$ only and causes the overall
redshift of the line. These curves correspond to the same values of
$\xi$ as in Fig.~\protect\ref{figline}.
\label{gfac}}
\end{figure}

The intrinsic shape of the iron line has been computed using a
Monte Carlo code {\sc{noar}} (Abrassart 2000b), 
coupled with the code
{\sc{titan}} (Dumont, Abrassart \& Collin 2000) computing the ionization
state of the gas for the assumed value of ionization parameter
$\xi{\equiv}F_{_{\rm{X}}}/n$ ($F_{_{\rm{X}}}$ denotes the incident X-ray
flux, $n$ is average number density of a cloud). The code {\sc{titan}}
serves to compute the radiative transfer in the Compton thick medium in
a broad-band frequency range, and it includes the computations of
opacity and temperature structure. It also gives local emissivity of the
gas in various components of the K${\alpha}$ line. The {\sc{noar}} code
uses the resulting opacity, it computes, in more detail, the transfer of
hard X-ray photons, and provides better description of the details of
hard X-ray spectra including Comptonization of the line within the hot
surface layers of the clouds. It also describes the energy deposit by
hard X-ray Compton heating for {\sc{titan}}. The two codes are thus
coupled and they both together offer the most accurate description of
the iron line emission from the irradiated medium, assuming it is
optically thick for electron scattering.

The computations were performed for a plane parallel slab corresponding
to a sphere with large radius, so that multiple reflections between the
clouds were neglected (they can effectively change only the value of the
ionization parameter).
The line profiles used in our computations of the cloud model are shown
in Figure~\ref{figline}. The pattern obtained for $\xi=10^2$ is centered
at 6.4\,keV, and it is so narrow that the line comes out almost
unresolved, and any smaller degree of ionization would result in even
narrower line. The line for $\xi=10^3$ shows the peak shifted to
6.7\,keV, and a broad red wing. The line for $\xi=10^4$ contains a
strong contribution of 6.9\,keV component arising from hydrogen-like
atoms; its red wing is extremely broad for such a high value of
ionization.

Since the numerical computations of intrinsic profiles are exceedingly
time consuming, we detach the line from its underlying reflected
continuum and discuss the line spectrum separately. Such a
simplification is possible because the line is not very sensitive to the
slope of the incident X-ray radiation, while the reflected continuum
depends on the slope considerably and would require computations of a
set of different incident continuum slopes.

Figure~\ref{gfac} shows how the intrinsic centroid energy of the line is
moved towards lower energy by factor $L(r)\,\gamma^{-1}(v_{\rm{k}}(r))$
in eq.~(\ref{g}). This term is function of radius only. Therefore the
graph shows roughly the mean energy of the expected observed line peaks
from Fig.~\ref{figline} depending on the radius of the clouds sphere.

In the next section we calculate the expected profiles for different
parameters of the model. In particular, we further discuss the effect of
broadening of the line, which arises from $\phi$-dependent term
$\bmath{v.s}$ in eq.~(\ref{g}) and results in large equivalent widths
and double-horn profiles in some cases.

\section{Results}
\label{results}

\begin{figure}
\epsfxsize=\hsize
\epsffile{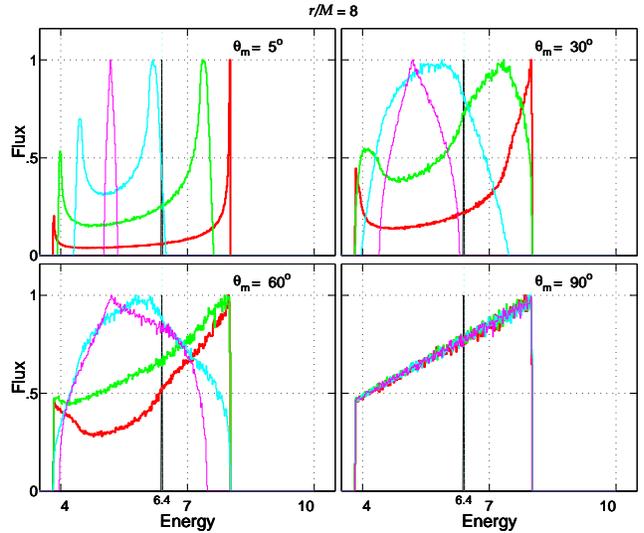}
\caption{The expected line profile produced by the clouds sphere with
$r=8M$. The intrinsic emissivity has been assumed as a delta line
at 6.4\,keV. The four panels correspond to different angular widths of
the clouds belt: $\theta_{\rm{m}}=5\dg$ (almost equatorial ring),
$30\dg$, $60\dg$, and $90\dg$ (strictly spherical distribution). In each
panel, four curves correspond to different values of the observer
inclination: $\theta_0=0$ (pole-on view, therefore the narrowest
profile), $30\dg$, $60\dg$, $90\dg$ (equator-on view, the broadest
profile). No obscuration is involved here. See the text for details.
\label{figt1}}
\end{figure}

\subsection{A narrow intrinsic line}
\label{delta}
The computations performed with the codes {\sc{titan}} and {\sc{noar}}
gave the Fe K${\alpha}$ line strong but practically unresolved for the
case of ionization parameter $\xi=100$ under the adopted energy
grid: $\Delta{E}\approx0.07$\,keV, and $\Delta{E}/E\approx0.01$. A faint
red wing is barely visible and it contains less than a few per cent of
the line flux. Therefore, in order to study the case of a neutral iron
line we could neglect the intrinsic width of the line and we performed
computations assuming that all photons were emitted with energy 6.4\,keV
(delta-type line with no continuum background). The observed line
profile is thus determined exclusively by radius of the clouds
distribution $r$, the level of departure of the clouds from the
equatorial plane $\theta_{\rm{m}}$, observer's angle $\theta_0$, and the
obscuration parameter $\omega$ which is further related to the covering
factor.

\begin{figure}
\epsfxsize=\hsize
\epsffile{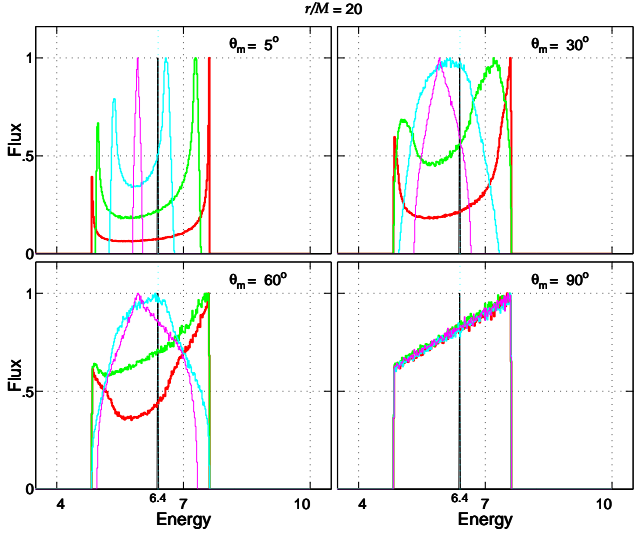}
\caption{The same as in Fig.~\protect\ref{figt1} but for $r=20M$. Such
intrinsically narrow profiles are expected in the case of low ionization
of the clouds. They can also serve as templates for convolution with
broader and more complex intrinsic profiles from a highly ionized medium.
\label{figt2}}
\end{figure}

\begin{figure}
\epsfxsize=\hsize
\centering
\epsffile{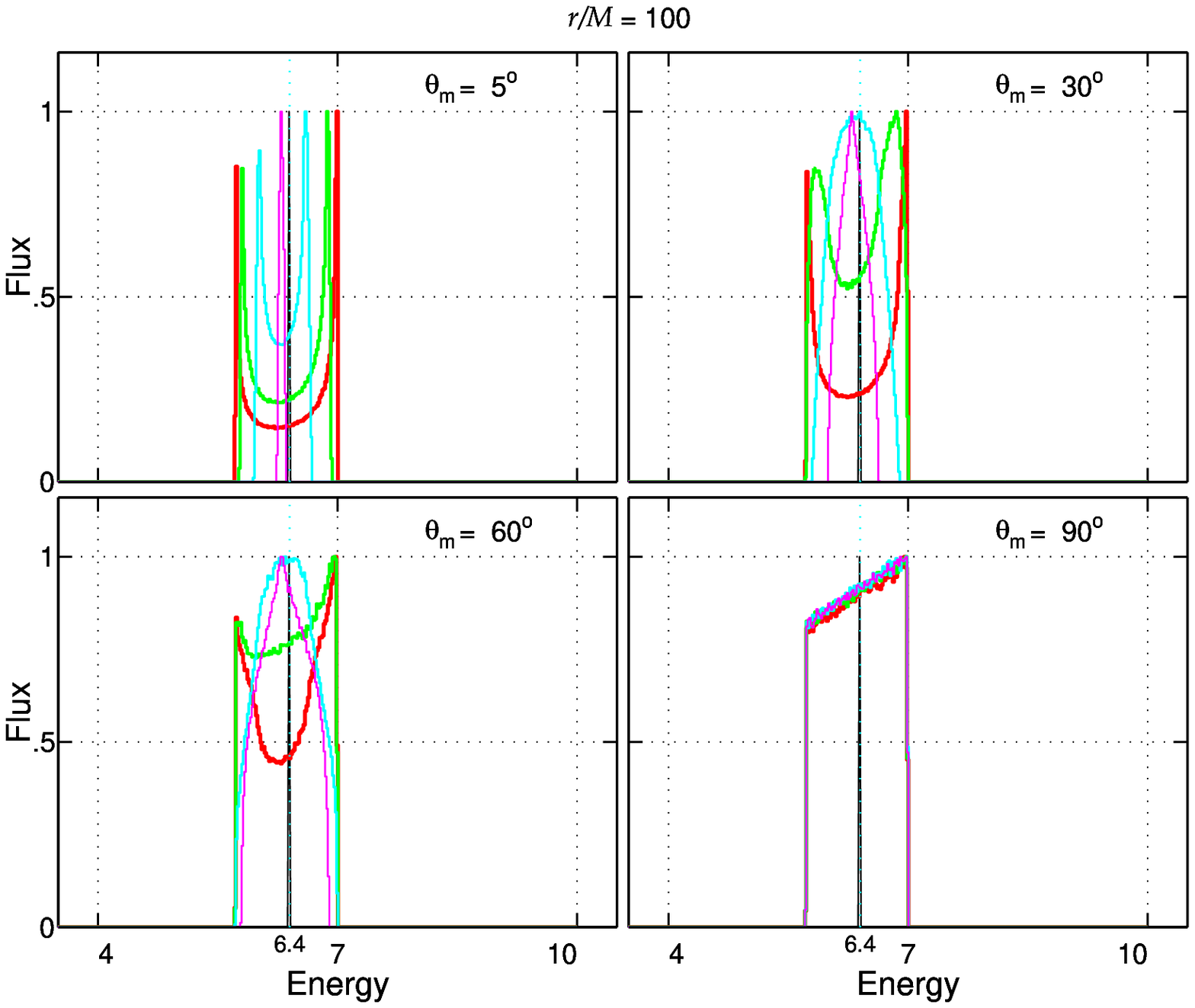}
\caption{The same as in Fig.~\protect\ref{figt1} but for $r=100M$.
\label{figt3}}
\end{figure}

\begin{figure}
\epsfxsize=\hsize
\epsffile{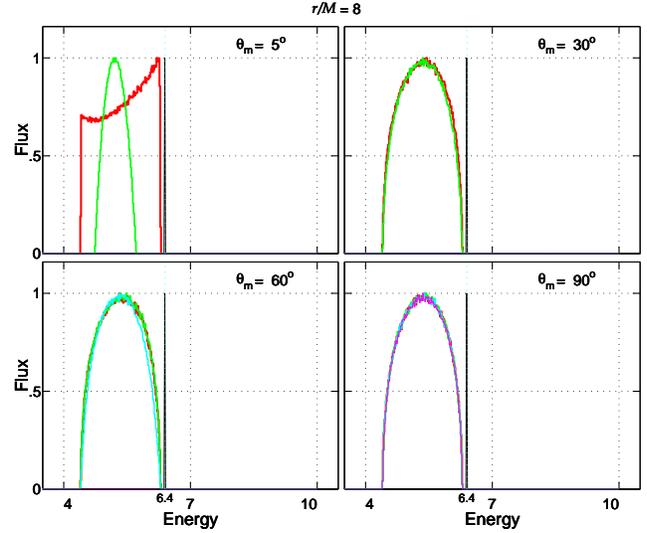}
\caption{The predicted spectral profiles corresponding to delta-type
line as in Fig.~\protect\ref{figt1}, but now the obscuration parameter
has been introduced with $\omega=30\dg$. The dependence of the profiles
on observer's inclination is substantially reduced in comparison with
the previous case of negligible obscuration. Here, $r=8M$.
\label{figp1}}
\end{figure}

\begin{figure}
\epsfxsize=\hsize
\epsffile{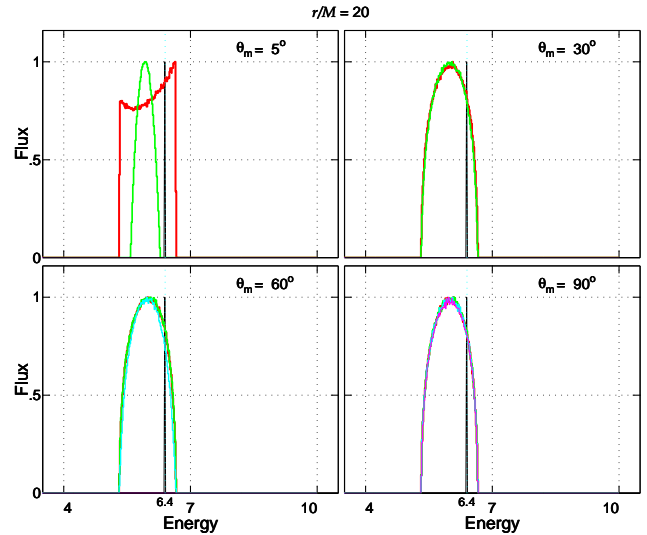}
\caption{The same as in Fig.~\protect\ref{figp1} but for $r=20M$.
\label{figp2}}
\end{figure}

We show the predicted profiles for three different radii $r$ of the
clouds distributions (Figures \ref{figt1}--\ref{figt3}). If the clouds
occupy a very narrow belt on the sphere ($\theta_{\rm{m}}\la5\dg$), the
resulting profiles are reduced to those obtained for a single ring
within the frame of the disc model of the line formation (Gerbal \&
Pelat 1981; Laor 1991). The corresponding values of $r$ and
$\theta_{\rm{m}}$ are given on top of each figure. Two horns of the line
are clearly seen with unequal heights due to the Doppler effect. The
dependence of the profiles on $\theta_0$ gets gradually diminished with
increasing sphericity of the clouds distribution, as can be checked by
comparing four panels with different $\theta_{\rm{m}}$. The profiles are
normalized to the maximum flux.

Figs.\ \ref{figt1}--\ref{figt3} correspond to zero obscuration of the
clouds ($\omega=90\dg$); the clouds do not shield each other and they
all are seen by the observer. We remark that the resulting profiles are
not completely smooth in these graphs because they are produced by
individual sources of light. Naturally, the curves would appear very
wiggly if the number of clouds and the corresponding covering factor
were small, but our simulations indicate quite large covering factors.

The resulting line profiles come out substantially different from the
previous case of zero obscuration if $\omega\ll\pi/2$: the profiles get
narrower, and they loose, partly or even completely, their double-horn
shapes, so characteristic for ring-type sources (Figures
\ref{figp1}--\ref{figp2}). Considered jointly with the underlying
continuum, there is less power in the line, so that equivalent width is
diminished. The dependence on observer's inclination is further reduced
by obscuration. One can easily deduce that the profiles do not depend on
$\theta_0$ if $\omega$ is small enough:
$\omega\leq\theta_0+\theta_{\rm{m}}-\pi/2$. Also, notice that the
constraints on the clouds visibility are not satisfied in case of very
small $\theta_0$, $\omega$ and $\theta_{\rm{m}}$, so that the curves are
shown only for appropriate combinations of these parameters.

What remains for the next section is to complete the above discussion by
computing the expected profiles of more complicated intrinsic
emissivities.
             
\subsection{A broad intrinsic line}
\label{intrinsic}
\begin{figure}
\epsfxsize=\hsize
\epsffile{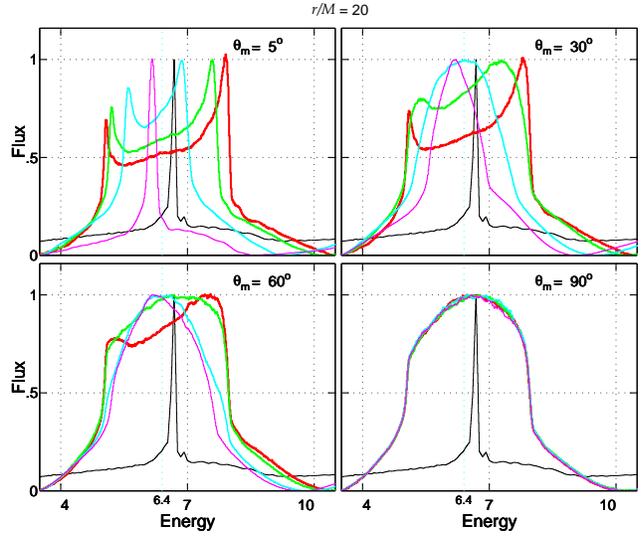}
\caption{The expected iron-line profiles (power-law continuum
subtracted) produced by the clouds at $r=20M$ for four values of the
inclination angle of observation $\theta_0$ ($0\dg$, $30\dg$, $60\dg$,
$90\dg$). Again, the panels are labeled with the adopted values of the
width of the belt, $\theta_{\rm{m}}$. The angle $\omega=90\dg$
corresponds to zero obscuration factor. The intrinsic spectrum is also
plotted with the maximum at $6.7\,$keV, corresponding to the assumed
ionization parameter, $\xi=10^3$.
\label{figli1}}
\end{figure}

\begin{figure}
\epsfxsize=\hsize
\epsffile{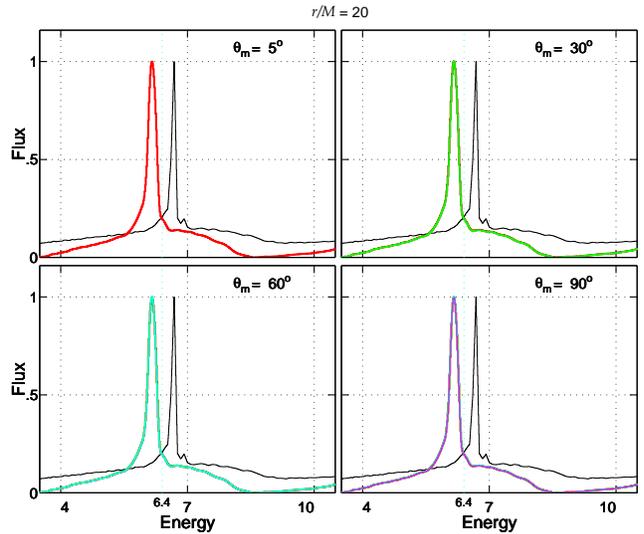}
\caption{The same as in Fig.~\protect\ref{figli1} but for $\omega=10\dg$.
\label{figli2}}
\end{figure}

Now we examine the spectral features with intrinsically broad profiles
and high ionization parameter ($\xi\ga10^3$; Abrassart 2000b). In each
case the continuum had been fitted by a power-law outside the feature
(4--9\,keV), and then the result was subtracted from the predicted
spectrum. In this way the expected spectral features have been obtained
(Figures \ref{figli1}--\ref{figli2}). We recall that the full (line and
continuum) intrinsic spectra are the result of radiation transfer around
6.4\,keV with multiple scattering among the clouds; they differ
substantially from those corresponding to simple delta-line profiles.
Here we present narrow band spectra; the actual energy range is shown on
abscissae. The two figures differ by obscuration parameter $\omega$:
obscuration was ignored in the former, while it was assumed to be
very large in the latter case.

Let us note that the values of $f_{\rm{c}}$ inferred from spectra
fitting procedures (and corresponding parameter $\omega$ of the model)
represent limiting boundary estimates of these quantities. One of the
reasons of uncertainty is the clouds velocity which until now has been
assumed fixed and equal to $v_{\rm{k}}(r)$ while, in case of strong
radiation pressure, the clouds motion may be non-Keplerian. The layer of
clouds can be partially supported by intercepted radiation flux
coming mainly from the center, so that steady-state motion of the clouds
becomes sub-Keplerian. Therefore, given the observed line width, the actual
covering factor will be less than that required on the basis of clouds
free orbital motion. In the case of quasi-spherical distribution, the
effect of radiation pressure is maximal for Eddington-type equilibrium
condition, $GMm_{\rm{p}}=\lambda$, where $m_{\rm{p}}$ is the proton
mass, and
\begin{equation}
\lambda=\frac{L\sigma_{_{\rm{T}}}}{4\pi(1-f_{\rm{c}})c}
\label{edd}
\end{equation}
is the outward-directed radiation support due to total 
luminosity $L$ corrected for partial obscuration. For large
covering factor the total luminosity is small ($f_{\rm{c}}\rightarrow1$,
$L\rightarrow0$) and the ratio on the right-hand side of eq.~(\ref{edd})
remains finite. Centroid energy of the line from such a static system of
the clouds is determined by the intrinsic spectrum and gravitational
redshift only, and the observed line width is reduced to the intrinsic
one.

\subsection{The case of MCG--6-30-15}
The Seyfert 1 galaxy MCG--6-30-15 is a notable example in which the
extensive observational material has been collected. It shows the
primary spectral slope of $\Gamma\approx2.00\pm0.05$, as determined on
the basis of joint {\it{ASCA}}/RXTE observations lasting
$\approx400$\,ksec with the cutoff energy at $\approx100$\,keV (Lee et
al.\ 1999).
The disc-line scheme has been rather successful also for this object
(Tanaka et al.\ 1995) and the conclusions drawn from the {\it{ASCA}\/}
data are in agreement with the analysis based on BeppoSAX observations
(Guainazzi et al.\ 1999). As a consequence, strong limits can be imposed
on alternative models (Fabian et al.\ 1995; Reynolds \& Wilms 2000).
However, in view of the complexity of the problem, it still seems
somewhat premature to condemn other models as nonviable. We also remark
that this object is probably somewhat exceptional case, though it is 
the best studied one at present.

\begin{figure}
\epsfxsize=\hsize
\epsffile{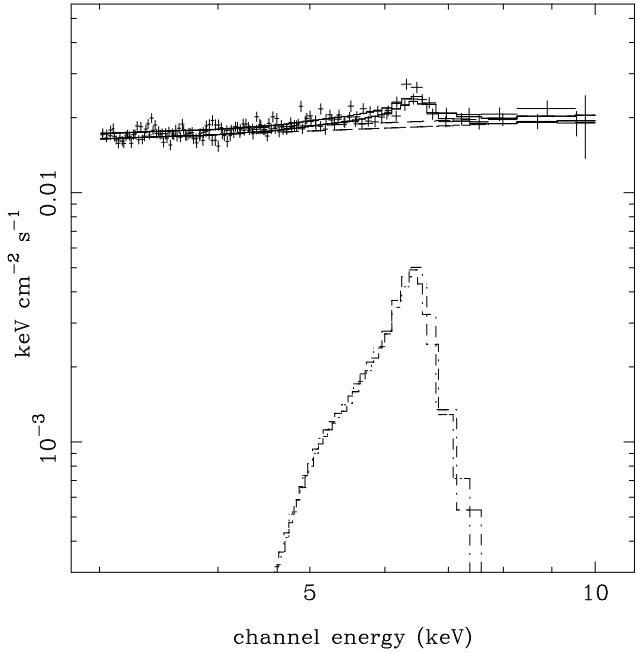}
\caption{The continuum and the Fe K$\alpha$ line profile from the model
 {\tt{cloudfe}} fitted to the {\it{ASCA}\/} data for MCG--6-30-15.
 Resulting parameter values are $r=29.1$, $\Gamma=1.856$. The remaining
 parameters were kept fixed: $\xi=10^4$, $\theta_{\rm{m}}=90\dg$,
 $\delphi=0.1$.
\label{xsp}}
\end{figure}

\begin{figure}
\epsfxsize=\hsize
\epsffile{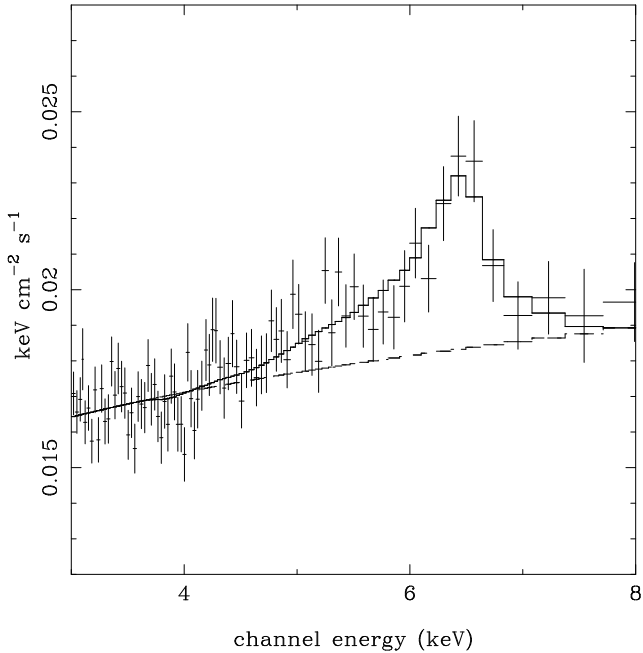}
\caption{A detail of the line and the best fit profile are shown
 with better resolution between 4--8\,keV.
\label{xl}}
\end{figure}

\begin{table*}
\caption{Spectral fitting of the model parameters to the data for
 MCG--6-30-15. Standard $\chi^2$ statistics was examined for different
 {\sc{xspec}} models. The first two lines in the table refer to standard
 models for the sake of comparison ({\tt{diskline}} and {\tt{powerlaw}}),
 while the rest corresponds to the user supplemented model
 ({\tt{cloudfe}}) as described in the text. The resulting values of
 three fitted parameters are given.
\label{tab:data}}
\begin{tabular}{ccrrrrr}
\hline\hline
Model                     & $\xi$  & Parameter \#1                    & 
Parameter \#2                   & Parameter \#3                       & 
$\chi^2$/d.o.f.\\
\hline
{\tt{diskline}}           & ---    & $\Gamma=1.810^{+0.019}_{-0.020}$ & 
$\beta=-2.45^{+0.34}_{-0.30}$   & $\theta_0=31.5^{+3.4}_{-3.2}$\,deg  & 
723.6/720\\
{\tt{powerlaw}}           & ---    & $\Gamma=1.723^{+0.020}_{-0.021}$ & 
---~~~                          & ---~~~                              & 
852.5/722\\
{\tt{cloudfe}}, spherical & $10^2$ & $\Gamma=1.728^{+0.021}_{-0.018}$ & 
$r=38^{+141}_{-37}\,GM/c^2$     & $\delphi=0.52^{+0.29}_{-0.39}$\,rad & 
804.1/720\\
{\tt{cloudfe}}, spherical & $10^3$ & $\Gamma=1.792^{+0.024}_{-0.022}$ & 
$r=48.4^{+10.2}_{-6.3}\,GM/c^2$ & $\delphi<0.19$\,rad                 & 
725.3/720\\
{\tt{cloudfe}}, ring-type & $10^3$ & $\Gamma=1.796^{+0.025}_{-0.022}$ & 
$r=49.0^{+10.6}_{-7.1}\,GM/c^2$ & $\theta_{\rm{m}}<21.7$\,deg         & 
724.9/720\\
{\tt{cloudfe}}, spherical & $10^4$ & $\Gamma=1.856^{+0.029}_{-0.029}$ & 
$r=29.1^{+3.7}_{-4.3}\,GM/c^2$  & $\delphi<0.13$\,rad                 & 
720.0/720\\
{\tt{cloudfe}}, ring-type & $10^4$ & $\Gamma=1.857^{+0.026}_{-0.024}$ & 
$r=29.8^{+5}_{-4}\,GM/c^2$      & $\theta_{\rm{m}}<19.4$\,deg         & 
721.6/720\\
\hline
\end{tabular}
\end{table*}

We tested our model of the line profile predicted by the clouds model by
fitting it to the {\it{ASCA}\/} data from July 1994 (Tanaka, Inoue \&
Holt 1994; Fabian et al.\ 1994; Tanaka et al.\ 1995; Ebisawa et al.\
1996). In order to avoid problems with modelling of the reflected
component, the contribution from the warm absorber (Otani et al.\ 1996),
and the cold absorption along the line of sight, we restrict our
discussion to the SIS data in 3--10\,keV band where the continuum can be
represented roughly by a single power law. Since the detailed fitting is
beyond the focus of the present work, we do not consider various
luminosity states separately (instead, we follow the initial approach of
Tanaka et al.\ 1995).

In order to have a reference for the results from our model we started
by fitting the {\it{ASCA}\/} spectrum with a power law and a disc-line
component (Fabian et al.\ 1989) from {\sc{xspec}} (Arnaud 1996); see
Table~\ref{tab:data} for the results of $\chi^2$ statistics. We assumed
the line energy at 6.4\,keV, consistently with the previous results. The
X-ray incident flux was allowed to vary with the distance as
${\propto}\,r^{-\beta}$ with inner and outer disc radii fixed at
$6\,GM/c^2$ and $10^3GM/c^2$ respectively. Then we compared those
results with the fits to our model in two different parameter ranges.

\subsubsection{The spherical configuration}
First we consider the case of spherically symmetric arrangement of the
clouds. The clouds cover the entire sphere
($\theta_{\rm{m}}=90\dg$), so that the inclination angle is of no
importance. Fitted parameters are: the photon index of the power law
$\Gamma$, radius of the clouds distribution $r$, and the visibility
factor $\delphi$ defined as
\begin{equation}
\delphi={\omega\over\pi/2}.
\end{equation}

Different versions of the model {\tt{cloudfe}} were examined using
{\sc{xspec}} supplemented by three local model subroutines. They differ
by corresponding values of ionization parameter:
\begin{description}
\item[(i)]~{an intrinsically narrow
line in the rest frame of the clouds; this case
refers to low ionization of the clouds surfaces where X-rays are
reprocessed, $\xi\la10^2$;}
\item[(ii)]~{a broad line with $\xi=10^3$ of the local emission;}
\item[(iii)]~{a broad line with $\xi=10^4$ of the local emission.}
\end{description}

The first of our models, with an intrinsically narrow line, does not
provide adequate representation of the data. Strong residuals remain (as
in the case of the Gaussian model), and the red wing cannot be reproduced
successfully. However, for intrinsically broad line (with
$10^3\la\xi\la10^4$) the red wing is formed locally as the result of
Comptonization, and it has roughly the required shape. This model
results in $\chi^2$ comparable to the {\tt{diskline}} profile
(Tab.~\ref{tab:data}). The larger value of the ionization parameter
gives marginally better fit than the lower one. Results of the fitting
procedure with SIS0 and SIS1 detectors are shown in Figure~\ref{xsp}.
The line is reproduced with better resolution in Figure~\ref{xl} (here
we show only SIS0 data for clarity).

Our calculation imposes limits both on radius of the clouds distribution
and on their covering factor. The former is constrained by requirement
of appropriate location of the peak of the line; the gravitational shift
has to balance somewhat increased energy of the line emission in this
model (Fig.~\ref{gfac}). The latter dependence reflects the fact that
small radius of the clouds distribution goes in hand with dramatically
strong smearing of the line, unless the effect of self-obscuration
prevents us from having a direct view of the clouds with the highest
velocities along the line of sight. We recall that the above discussion
assumes Keplerian motion of the clouds; numerical values given in the
column $\delphi$ of Tab.~\ref{tab:data} must be increased (less
obscuration) if sub-Keplerian motion of the clouds is taken into
account, as discussed in previous section.

\subsubsection{The ring-type configuration}
In order to see whether we can detract the strong requirements on the
clouds obscuration at the expense of introducing partially flattened,
ring-type geometry, we fitted the model assuming $\delphi=0.5$ and
varying $\theta_{\rm{m}}$. To illustrate the procedure, let us fix
inclination angle at $\theta_0=30\dg$, and consider two values of the
ionization parameter, $\xi=10^3$ and $\xi=10^4$. The results are
referred as ring-type models {\tt{cloudfe}} in Tab.~\ref{tab:data}. The
quality of the fits is comparable to those obtained for spherically
symmetric case, however, the configuration comes out very flat, almost a
ring indeed. We could not fit the data assuming $\delphi\rightarrow1$
(the case of negligible obscuration) because such a flat distribution of
the clouds clearly requires a substantial range of radii contributing to
the spectrum (rather than a narrow ring). This conclusion is valid
independently from the intrinsic broadening of the line due to internal
Comptonization.

On the basis of several other attempts to find acceptable fits with
different obscuration, we reached the conclusion that the currently
available data is still not good enough to distinguish definitively in
this model between a flat geometry and a spherically symmetric geometry
of the accretion flow.

\section{Discussion}

The effect of obscuration between the clouds plays an important role 
in the model: it restricts line-of-sight velocities of the clouds and
suppresses the dependence of observed line profiles on inclination in
comparison with the standard formulation of the disc-line model. 
Another interesting attribute of
the clouds model is the requirement of high ionization of the material 
which is
responsible for the iron-line emission: $\xi\ga10^3$. Such a high
ionization has two consequences which we can briefly outline.

\subsection{The role of high ionization}
Since the red wing of the line is mostly due to Comptonization
within the clouds surface layers (instead of kinematics of the accretion
disc), the observed variability of the line profile must be also linked
with the changes of the physical state of the clouds surfaces (Ebisawa
et al.\ 1996; Lee et al.\ 1999). Self-consistent calculations of the
resulting spectra present a difficult task because the thermal stability
of the gas irradiated by hard X-ray photons depends on many factors at
temperatures $10^5$--$10^6$\,K (Krolik 1999; R\'{o}\.{z}a\'{n}ska 1999,
and references therein). Even a minor change in the incident radiation
may cause a dramatic response of the temperature and, consequently, the
density of the irradiated gas through its expansion/contraction under
constant pressure. Such fluctuations mean a significant change of the
ionization parameter and of the intrinsic shape of the emitted line,
without major alteration of the total luminosity of the source or the
total flux in the iron line.

The demand of high ionization of the medium responsible for the line
emission makes a difference from almost neutral environment of
illuminated disc-type accretion flows, and there is hope to find its
imprint also in the continuum. However, the limited spectral range of
{\it{ASCA}\/} does not allow us to exploit such a possibility and to
differentiate between the disc and the clouds models.

Also, we cannot presently distinguish the spherical configuration with
strong self-obscuration of the clouds from the case of much flatter
geometry and less obscuration. A better spectral resolution is needed in
order to make the X-ray spectroscopy really a powerful tool and to
determine the form of the accretion flow. Further work on the models is
also needed to demonstrate true differences between the models.

\subsection{The role of hydrostatic equilibrium}
The calculations of local emissivity (Sect.~\ref{secemis})
have been carried out under constant density approximation.
If the clouds have enough time to achieve the hydrostatic equilibrium, the 
radiative transfer should be performed assuming constant pressure instead
of constant density within the reprocessing medium. Computations under the
constant pressure assumption or, more generally, under the assumption of
hydrostatic equilibrium show a specific temperature profile due to the
radiative thermal instability at the intermediate temperatures (see e.g.\
Krolik, McKee \& Tarter 1981; R\'o\.za\'nska \& Czerny 1996;
R\'o\.za\'nska 1999; Nayakshin et al.\ 1999). The irradiated medium is at
the inverse-Compton temperature in the upper layers while being cold inside,
with rather sharp transition determined by the effect of heat conduction
(R\'o\.za\'nska 1999). Nayakshin et al.\ (1999) argue that the intermediate
stable branch at the temperature about $10^6$\,K has a negligible influence
on the formation of spectral features, while preliminary results of 
Monte Carlo simulations (\.Zycki et al., in preparation) indicate that 
a possible enhancement of the iron abundance (crucial for the development 
of this
branch; Lee et al.\ 1999) by a factor of $\sim 2$ may change this conclusion. 

The problem remains unsettled for the moment, however, if the
inverse-Compton temperature itself is low enough to allow for the
formation of the K$\alpha$ line, then the line forms in a 
broad zone and the results of the
computations do not depend crucially on the assumed density profile.
This is quite possibly the case for the source MCG--6-30-15.

\subsection{The role of inverse-Compton temperature}
The spectral shape in the hard X-ray band was well determined 
for this object from the SAX
data (Guainazzi et al.\ 1999). The best fit model gave the photon
index $\Gamma=2.04$ and the high-energy cut-off at 130\,keV. The
inverse-Compton temperature corresponding to such a spectrum is equal to
$2.02 \times 10^7$~K, however, the soft photons emitted by   
the clouds and observed as the Big Blue Bump component reduce this
value further.

The object is highly reddened in the optical/UV band. The extinction
estimate
based on the Balmer line indicates the $E$(B$-$V) value in the range of
0.61--1.09 (Reynolds et al.\ 1997). Assuming a minimum value for the 
extinction these authors determine $L$(NIR$-$UV)$=0.3\,L$(X). 
Somewhat larger values are favoured because too low bolometric
luminosity of the internal source makes a large far-IR luminosity of this
object difficult to explain. The upper limit on the extinction rises the flux
at $\log(\nu)=15$ by a factor of 12 in comparison with the lower limit.
This fact together with an unconstraint high-frequency extension of the Big
Blue Bump introduce significant uncertainty in the shape of the broad-band     
spectrum. Therefore, we can assume a standard shape of the spectrum, as in
Abrassart \& Czerny (2000), i.e.\ $L$(NIR$-$UV)$=3\,L$(X). The observed
ratio is slightly different from the intrinsic spectrum seen by the clouds
(Abrassart \& Czerny 2000; see their Eq.~(12)). The correcting
factor to this ratio is typically equal to 1.05--1.11. 
Assuming the value of 1.1 we
finally obtain the Compton temperature of  $4.7\times10^6\,$K as a
characteristic value for MCG--6-30-15.

The derived value falls just within the temperature range which resulted from
our computations. The upper layer of the clouds, down to $\tau_{\rm{es}}=1$,
has the temperature of $6.5\times10^6\,$K assuming the
parameter $\xi=10^4$, and $1.8\times10^6\,$K for $\xi=10^3$.
Therefore, in the case of the presented computations, the presence or
the absence     
of hydrostatic equilibrium does not seem to pose a crucial question.
This is quite fortunate because, under our present insufficient understanding
of the clouds evolution, it is by no means clear which of the two
approximations is more appropriate to the clouds that we consider -- whether 
it is constant density or rather constant pressure assumption.
Notice that for other Seyfert galaxies than MCG--6-30-15
most probably their inverse-Compton temperatures will
tend to be higher because their spectra are harder.

\section{Conclusions}
We presented a simple cloud model of AGN in which self-consistent
computations of the intrinsic X-ray emission are supplemented by the
phenomenological description of the clouds distribution. We demonstrated
that satisfactory fits to the iron K${\alpha}$ line feature can be
obtained, and the model should be thus considered as viable and
developed further, so that the phenomenological parameters are derived from
the physical model of the clouds origin. We envisage the origin of the
clouds due to erosion of the inner accretion disc. Detailed physics of
the clouds formation remains beyond the scope of this paper but the
adopted parameters ($r$, $\theta_0$, $\theta_{\rm{m}}$, $\omega$) should
reflect this scheme in a natural manner.

We remark that the scheme described in the present model is not
restricted by recent considerations of Reynolds \& Wilms (2000) because
our model falls into that category of models in which the X-ray source
is not viewed through the Comptonizing medium, and the clouds experience
somewhat different continuum than the observer (Abrassart \& Czerny
2000).

Observed profiles must be influenced by several other effects which were
neglected in this paper due to our ignorance of the detailed model of
the clouds properties. One can however anticipate the expected
implications of a more refined model.

First, we ignored gravitational lensing for which one can expect a
similar conclusion like for partial obscuration and sub-Keplerian motion
of the clouds: the lensing effect enhances radiation coming from the
clouds near the line of sight and on the remote side of the sphere. If
included, the effect decreases the covering factor derived here by
fitting the line profiles.

Second, we did not consider radial distribution of the clouds. This
could be introduced in a straightforward manner but the assumption of
almost constant radius provides a very good approximation in the current
model. Hot parts of the clouds surfaces are those which lie innermost,
where they are subject to strong primary irradiation by X-rays. The
impact of the radial distribution can be absorbed in parameter $\omega$
(enhanced obscuration).

We conclude by emphasizing the fact that the present model differs from
the disc-line model in the mechanism how the intrinsic spectrum is
produced. There are however also common aspects of the two schemes.
Indeed, a ring-like distribution of the clouds can be treated as a
special case (small $\theta_{\rm{m}}$) resulting in profiles comparable
to the disc-line model with appropriate emissivity. Notice, however,
that here we obtained reasonable fits under the assumption of spherical
or almost spherical clouds distribution. Certain degree of the
flattening (characterized here by $\theta_{\rm{m}}$) is natural to this
model where the clouds origin is in continuous erosion of the inner
disc, but, on the other hand, $\theta_{\rm{m}}$ cannot be too small
because substantial Componization requires large covering factor.
Finally, we remark that successful fits require strong gravitational
field of the central body which shifts the line centroid towards lower
energy.

\section*{Acknowledgements}
The authors benefited from numerous discussions with Suzy Collin and 
Anne-Marie Dumont. They are grateful also to Piotr \.{Z}ycki for help with 
the {\it{ASCA}\/} data, to Giorgio Matt for useful remarks about
fitting the iron line, and to the referee for pointing out the question
of hydrostatic equilibrium of the clouds and several other comments which
helped us to improve our discussion. This research has made use of the TARTARUS
database which is supported by Jane Turner and Kirpal Nandra under NASA
grants NAG5-7385 and NAG5-7067. Part of this work was supported by the
grant 2P03D01816 of the Polish State Committee for Scientific Research, 
and by Jumelage/CNRS No.~16 ``Astronomie France/Pologne''. V\,K
acknowledges hospitality of the Observatoire de Paris-Meudon, and
partial support from the grants GAUK 63/98, GACR 202/\lb{2}98/\lb{2}0522
and 205/\lb{2}00/\lb{2}1685 in the Czech Republic.

\label{lastpage}

\end{document}